# MacLaurin and Morality

Isobel Falconer, School of Mathematics and Statistics, University of St Andrews, Scotland

David Horowitz, Golden West College (Emeritus), Huntington Beach, CA

## 1. Introduction

Scottish mathematician Colin MacLaurin (1698-1746) is best known for his *Treatise of Fluxions* (1742), and the eponymous power series. Little known is that in 1714, aged sixteen, MacLaurin wrote a short manuscript attempting to apply Newtonian principles to morality. *De Viribus Mentium Bonipetis* (On the good-seeking forces of minds) remained hidden in the Colin Campbell Collection at the University of Edinburgh for over 250 years;[1] it was only uncovered in the late twentieth century. *De Viribus* provides a remarkable glimpse into how the young MacLaurin dealt with early Newtonianism, the tenets of the Church of Scotland, and the nascent interface between science and religion just prior to the dawn of the Scottish Enlightenment.

Perhaps the most intriguing aspect of *De Viribus* is the reflections on Scottish Presbyterian morality that MacLaurin interjects throughout his mathematical discussion. These are often oblique, and one must look to his mathematics, his contemporaries, and surrounding social context, to understand them. In the process, one gains insight into both MacLaurin's background and religious thought, and the nature of teaching in Scottish universities at the time.

This chapter examines the content of *De Viribus Mentium Bonipetis* in depth for the first time, drawing on Ian Tweddle's translation.[2] After brief a brief biographical background, it will discuss the manuscript and its history, and the intellectual climate within which MacLaurin was working, before moving on to interpret the mathematics and theology of *De Viribus*.

---

[1] Colin Maclaurin, *De Viribus Mentium Bonipetis*, Colin Campbell Collection, Edinburgh University Library Centre for Research Collections (E.U.L.) MS 3099.15.8, [1714] repr. in Ian Tweddle, 'An early manuscript of MacLaurin's: mathematical modelling of the forces of good; some remarks on fluids', 2008 rev. 2018 <https://mathshistory.st-andrews.ac.uk/Publications/tweddle_maclaurin.pdf> [accessed 25 September 2025].

[2] Tweddle, 'An early manuscript'. Throughout this chapter, references to specific pages in *De Viribus* will be given as folio number in the manuscript followed by page number in Tweddle's translation, e.g.(fol. 5$^v$, Tweddle p. 5). A preliminary version of this chapter, containing additional detail, especially of theological points, is available online: David Horowitz and Isobel Falconer, 'MacLaurin and Morality', https://thecolinmaclaurin.wordpress.com/ [accessed 25 September 2025].

Interpreting *De Viribus* in this way provides a lens for examining the intellectual context of early eighteenth-century Scotland, within which such an endeavor made sense to the youthful MacLaurin, newly graduated from Glasgow University. *De Viribus* not only demonstrates that ideas of mathematising morality were being canvassed in Scotland much earlier than the better known, but less sophisticated, attempts by Francis Hutcheson, David Hume and George Turnbull, but suggests continuities with approaches to mathematization in earlier centuries, providing a bridge into the Enlightenment era.

## 2. Biographical Background

Colin MacLaurin was born in February 1698 in the parish of Kilmodan in western Scotland, the third child of the local minister, who died six weeks after MacLaurin's birth. He was raised by his mother, Mary Cameron, and following her death in 1707, by his uncle, the Reverend Daniel MacLaurin (d. 1720), in Kilfinan in Argyllshire.[3] At the time of his birth, MacLaurin's family was firmly established in the ecclesiastical life of Scotland, having benefited from the rise to power of the Presbyterian, more Calvinist-leaning, faction in the Church of Scotland following the accession of William and Mary to the British throne.[4] Colin and his elder brother, John, had strong religious upbringings.[5]

In 1709 Colin MacLaurin enrolled in the four-year masters degree at the University of Glasgow.[6] At eleven, he was not unusually young in this period. The university operated under the traditional Scottish regent system wherein each cohort of students was overseen by a single generalist supervisor (regent) who gave their lectures in all subjects throughout the degree course.[7] In a variation that presaged a move to a specialist professorial system, Glasgow had one regent, the Professor of Greek, dedicated to first year students; the students then passed to another regent who taught their second and subsequent years.[8] MacLaurin is listed among fifteen first-year students supervised by Alexander Dunlop (1684-1747).[9]

---

[3] Fuller biographies are: Colin MacLaurin and Patrick Murdoch, *An Account of Sir Isaac Newton's Philosophical Discoveries…*, (London: the Author's children, 1748) pp. i-xx; John Smith, *Iconographia Scotia: or Portraits of Illustrious Persons of Scotland, Engraved from the Most Authentic Paintings &c*. (London: Robert Wilkinson, [1798?]), fols 44$^v$-48$^r$; Charles Tweedie, 'A study of the life and writings of Colin Maclaurin', *The Mathematical Gazette,* 8 (1915-1916), pp. 133-151; 9 (1917-1919), pp. 303-306; 10 (1920-1921), p. 209; G.A. Gibson, 'Sketch of the history of mathematics in Scotland to the end of the I8th century: Part II', *Proceedings of the Edinburgh Mathematical Society.* 1 (1928), pp. 71-93; Robert Schlapp, 'Colin Maclaurin: a biographical note', *Edinburgh Mathematical Notes*, 37 (1949), pp. 1-6; Herbert Westren Turnbull, *Bi-centenary of the Death of Colin Maclaurin (1698-1746),* (Aberdeen: University Press, 1951); Johnston R. McKay, 'The MacLaurin papers', *The College Courant, The Journal of the Glasgow University Graduates Association,* 25 (1973), pp. 30-34; Stella Mills (ed.), *The Collected Letters of Colin MacLaurin* (Nantwich: Shiva, 1982) pp. xv-xx; Erik Lars Sageng, 'MacLaurin, Colin (1698-1746)', in *Oxford Dictionary of National Biography*, 2004.

[4] Mills, *Collected Letters*, p. xv; Stewart J. Brown, 'Religion and Society to c.1900', in *The Oxford Handbook of Modern Scottish History*, ed. by T. M. Devine and Jenny Wormald (Oxford University Press, 2012), doi:10.1093/oxfordhb/9780199563692.013.0005.

[5] John MacLaurin (1693-1754) became a leading Church of Scotland theologian in Glasgow. See Richard Sher, 'Maclaurin, John (1693-1754)', in *Oxford Dictionary of National Biography*, 2004.

[6] The MA was (and is) the first degree at Scottish universities, although the BSc is now awarded in sciences; some of the newer universities (e.g. University of Stirling) have begun to award the BA degree.

The regent for MacLaurin's, and the rest of his cohort's, final three years was the polymath Scottish philosopher Gershom Carmichael (1672-1729), a leading figure in the early Scottish Enlightenment.[10] Carmichael's description of his teaching duties at Glasgow outline the curriculum MacLaurin would have studied:

> In the Magistrand [fourth] Year (if God spare them & me together), ye first work must be to compleat what yet remains undone of ye Ethicks, & then, if possible, again to glance over ye Pneumatick & Ethick Theses:[11] tho' at the same time, I must endeavour, with all convenient Speed, to get them thro', at least, ye sixth Book of Pardies Elements [of Geometry],[12] without which they can make verry few Steps to purpose in ye Physicks: I must likewise give them a touche of some other parts of Geometry, as time will allow; but for their more thorow acquaintance with them, refer them to ye Professor of Mathematicks.[13]

Thus, the study of mathematics beyond basic geometry was an extracurricular option. For MacLaurin, the 'Professor of Mathematicks' was Robert Simson, who had arrived at Glasgow in 1711 and held the chair for fifty years, becoming a major figure in Scottish mathematics.[14] There is no record of MacLaurin having formally studied with Simson, although historians frequently assume that he did; the pair were certainly friends in later years.[15] A recently-rediscovered notebook of MacLaurin's may provide additional evidence.[16]

---

[7] Christine Shepherd, 'Philosophy and science in the arts curriculum of the Scottish universities in the 17th century', (University of Edinburgh Ph. D. thesis, 1974), pp. 18-60.
[8] For the Scottish curriculum, see Eric Forbes, 'Philosophy and science teaching in the seventeenth century', in *Four Centuries: Edinburgh University Life 1583-1983*, ed. by Gordon Donaldson, (University of Edinburgh, 1983), p. 30. For the spread of scientific ideas across the Scottish universities see: John L. Russell, 'Cosmological Teaching in the Seventeenth-century Scottish Universities', *Journal for the History of Astronomy*, Part 1, 5.2 (1974), pp. 122-132; Part 2, 5.3 (1974), pp. 145-154; Christine M. Shepherd, 'Newtonianism in Scottish universities in the seventeenth century', in *The Origins and Nature of the Scottish Enlightenment*, ed. by R. H. Campbell and Andrew S. Skinner (John Donald, 1982), pp. 65-85.
[9] University of Glasgow, *Munimenta Alme Universitatis Glasguensis: Records of the University of Glasgow: from its Foundation till 1727*, ed. by Cosmo Innes, 4 vols, (Glasgow: [Maitland Club], 1853), III p. 195.
[10] Shepherd, 'Philosophy and science…' p. 376; for accounts of Carmichael see Gershom Carmichael, *Natural Rights on the Threshold of the Scottish Enlightenment*, ed. By James Moore and Michael Silverthorne, trans. by Michael Silverthorne, (Liberty Fund, 2002), pp. 377-388); ——, 'Carmichael, Gershom (1672-1729)', in *Oxford Dictionary of National Biography*, 2003; Mark H. Waymack, 'Moral Philosophy and Newtonianism in the Scottish Enlightenment' (Johns Hopkins University Ph.D. thesis, 1986).
[11] Pneumatick was the study of spiritual being, including the human soul – and related to ethics and questions of free will and natural or Divine law.
[12] An introductory Euclidean geometry by Ignace Gaston Pardies (1636-1673), first published in French in 1671. A copy of the 1694 Latin edition, *Elementa geometriæ*, is in Glasgow University Library, which may be the edition used by Carmichael <https://datb.cerl.org/estc/R40057> [accessed 25 September 2025].
[13] Carmichael, *Natural Rights*, p. 379.
[14] E. I. Carlyle, rev. by Ian Tweddle, 'Simson, Robert (1687-1768)' in *Oxford Dictionary of National Biography*, 2004.

MacLaurin received his MA degree on 23 June 1713 for his thesis *De Gravitate, aliisque Viribus naturalibus* (On gravity, and other natural forces).[17] Maclaurin chose this topic himself. The previous system, whereby regents set thesis titles, was just phasing out, and Maclaurin's, along with two other students', were the first individual theses.[18] Besides discussing its physical manifestations, MacLaurin asserted in his thesis that the importance and ubiquity of gravity were manifest verifications of the existence of a divine Creator.[19] Despite the apparent Newtonian origin of MacLaurin's thesis, it contained no explicit mathematics and no mention of fluxions or fluents, which reinforces the conclusion that Simson's teaching was peripheral to the compulsory curriculum and was not needed to satisfy the requirements for an MA.

On completing his MA, Colin MacLaurin began studying Divinity at Glasgow, and probably continued a friendship with Simson, learning more mathematics. But after a few months he withdrew and returned to live with his uncle Daniel in Kilfinan. Sageng has suggested that he was disgusted with the theological controversies besetting the Church of Scotland.[20] However, a related reason may be that the Patronage Act of 1712 had restored the power of appointing local ministers from parishioners to the traditional lay patrons;[21] hence securing a patron became more important than a university qualification in becoming a Church of Scotland minister.

Daniel MacLaurin may have been looking out for his nephew's future when he sent a copy of Colin's thesis to the Reverend Colin Campbell of Achnaba (1644-1726) in 1713, asking for the clergyman's opinion of the work: 'I send it to you, for you know some little of the Mathematicks, I beg the favour of your remarks'.[22]

---

[15] e.g. Sageng, 'Maclaurin, Colin'. MacLaurin is conspicuously absent in Trail's list of distinguished students of Robert Simson at Glasgow, and from Brougham's later list. This may indicate that any teacher-pupil relationship was informal. See William Trail, *Account of the Life and Writings of Robert Simson, M.D.* (London: G. & W. Nicol, 1812), p. 5-6; Henry Brougham, *Lives of Men of Letters and Science,* ( Charles Knight, 1845), p. 482.
[16] Colin MacLaurin, 'Student Notebook 1712-1714'. University of Glasgow Archives & Special Collections (GUA) GB 247, ASC 010; David Aaron Horowitz, 'The circuitous journey of the MacLaurin student notebook', *Journal of the Edinburgh Bibliographical Society*, 20 (2025), 45-5. Among other things, the notebook contains the student dictats that MacLaurin took during Carmichael's lectures; extracts are translated in Shepherd, 'Philosophy and science…'.
[17] University of Glasgow, *Munimenta,* p. 48; Colin MacLaurin, *Dissertatio Philosophica Inauguralis, de Gravitate, aliisque viribus Naturalibus*, (Edinburgh: Apud Robertum Freebairn, Typographum Regium, 1713) repr. and trans. in Ian Tweddle, *MacLaurin's Physical Dissertations* (Springer, 2007).
[18] Forbes, *Philosophy and Science Teaching* p. 29; 'Minutes of the Glasgow Faculty Meeting of 20 May 1713', G.U.A. 26631.
[19] MacLaurin, *Dissertatio De Gravitate* p. 4; trans. Tweddle, *MacLaurin's Physical Dissertations*, p. 18.
[20] Erik Lars Sageng, 'Colin MacLaurin and the Foundations of the Method of Fluxions' (Princeton University Ph.D. thesis, 1989), p10.
[21] Thomas Ahnert, *The Moral Culture of the Scottish Enlightenment, 1690-1805* (Yale University Press, 2014).
[22] Daniel Maclaurin, 'Letter to Colin Campbell of Achnaba, 1713', E.U.L. ms 3099.15.4.

Campbell, a Presbyterian-leaning minister, and a mathematician in his own right, was an ideal recipient of Daniel MacLaurin's message. Like Marin Mersenne in France and Henry Oldenburg in England, Campbell functioned as a sort of clearinghouse where scholars could send their work for review and dissemination.[23] He responded favorably to Daniel MacLaurin in an undated letter, admiring how the young MacLaurin, 'in the age of 15, should be so well versed in the true Principles of the Physicks'.[24]

Encouraged, in September 1714, Colin MacLaurin wrote directly to Campbell, enclosing an essay that he had written during the previous year — *De Viribus Mentium Bonipetis* (On the Good-seeking Forces of Minds) — the subject of this chapter. As he made clear to Campbell, it is an essay in which principles of early Newtonianism are applied to the study of Christian morality.

> But since I have mentioned the use of Mathematics I shall beg your pardon for troubling you with some thoughts I have relating to their use in morality. I have sent them to you under the title of De viribus mentium Bonipetis.[25]

This is the only reference to *De Viribus* in the letter. There is no record of Campbell's reaction to *De Viribus*, and MacLaurin never mentioned it again.[26] His short essay lay dormant in the Colin Campbell papers for nearly 250 years.

## 3. Considerations of the Manuscript

Until the late-twentieth century, *De Viribus* went virtually unnoticed. It was never mentioned in any of the major biographies of MacLaurin.[27] It was unearthed in the 1950s by John Carnegie Eaton (1915-1972) who prepared a partial transcription, but that, too, lay dormant.[28] Finally, in the mid-1980's Erik Sageng brought *De Viribus* to light in his 1989 doctoral thesis. He commented:

> It is easy to smile at De Viribus Mentium Bonipetis, but the graphing of qualities had been a standard part of the scholastic curriculum since its introduction by Nicole Oresme in the mid-fourteenth century, and more sophisticated moral

---

[23] John Henry, 'Campbell, Colin, of Achnaba (1644-1726)', in *Oxford Dictionary of National Biography*, 2004: Phillippe Bernhard Schmid, 'The Workplace of Enlightenment: Colin Campbell and the Repurposing of Paper', *Journal for Eighteenth-Century Studies*, 48.2 (2025), pp. 119–47.
[24] Colin Campbell, 'Letter to Daniel Maclaurin',(E.U.L. ms 3099.15.7).
[25] Colin Maclaurin, 'Letter to Colin Campbell, 12 September 1714', E.U.L. MS 3099.15.1. See also Mills, *Collected Letters*, [Letter 116].
[26] Six years later MacLaurin wrote to Campbell, 'I know not If [sic] you received the letter I sent you in 1714 for I have never heard from you […]' (Colin Maclaurin, 'Letter to Colin Campbell, 6 July 1720', E.U.L. MS 3099.15.2; see also Mills, *Collected Letters*, [Letter 117]).
[27] See fn.3.
[28] John Carnegie Eaton, transcript of *De Viribus,* G.U.L. MS Gen. 1332, Box 1. For information on Eaton see McKay, 'MacLaurin papers' and Mills, *Collected Letters,* p. xi.

philosophers than MacLaurin tried their hand at the quantification of various sorts of 'goods.'[29]

Since Sageng, a few others have glanced at *De Viribus* and offered insights.[30] However, it was Ian Tweddle who first carefully analyzed MacLaurin's manuscript, translating *De Viribus* and providing notes on its technical content.[31] Tweddle approached it as a mathematical document and did not examine the historical or personal context in which Maclaurin composed it.

The manuscript is a 17-page booklet in MacLaurin's handwriting, measuring about 9 cm by 15 cm. *De Viribus Mentium Bonipetis* (On the Good-seeking Forces of Minds) comprises the first ten (numbered) pages; the final seven (unnumbered) pages are titled *Prop. altera.*[32] In *Prop. altera* MacLaurin, without referring explicitly to Newton, applied fluxional calculus to prove propositions 21 and 22 on hydrostatics of Book II of the *Principia,* which Newton had left unproven.

*De Viribus* applies fluxional calculus to the question of the nature of the supposed attractive force between 'good things' and the mind. Throughout, the testbed for his mathematical models is the early eighteenth-century Scottish religious mores which permeated Maclaurin's life and thought. In his translation, Tweddle passed over the often cryptic statements about morality, and Grabiner whimsically commented:

> A sort of trial run of the Newtonian style was Maclaurin's youthful attempt […] to build a calculus-based mathematical model for ethics. In a Latin essay still (perhaps mercifully) unpublished today, "De Viribus Mentium Bonipetis" ("On the Good-Seeking Forces of Mind [sic]"), Maclaurin mathematically analyzed forces by which our minds are attracted to different morally good things.[33]

Yet there is more to *De Viribus* than a misguided early exercise in calculus. One can interpret it as an attempt to reconcile the precepts of his religion with his academic background. It provides a valuable insight into the intellectual climate within which such an endeavor made sense — an era in which the potentially applicable domain of Newtonian calculus was still fluid and, for many, the evidence of Scripture took precedence over that of the natural world. To make sense of this interplay of mathematics and theology, we first look at the early 18th century Scottish context within which Maclaurin was writing.

---

[29] Sageng, 'Colin Maclaurin' PhD. thesis, pp. 126-127.
[30] Harro Maas, 'Where mechanism ends: Thomas Reid on the moral and the animal oeconomy', *History of Political Economy* 35 (annual supplement 2003), pp. 338-360; Judith Grabiner, 'Newton, Maclaurin, and the authority of mathematics', *The American Mathematical Monthly* 111 (2004), pp. 841-852; ——, 'The role of mathematics in liberal arts education', in *International Handbook of Research in History, Philosophy and Science Teaching*, ed. by Michael R. Matthews, (Springer, 2014), pp. 793-836; Christian Constanda, *Dude, Can You Count? Stories, Challenges, and Adventures in Mathematics* (Springer, 2009); Olivier Bruneau, *Colin Maclaurin ou L'obstination Mathématicienne d'un Newtonien* (Presses Universitaires de Nancy, 2011); —— 'Colin Maclaurin (1698-1746): a Newtonian between theory and practice', *British Journal for the History of Mathematics* 35 (2020), pp. 52-62; Paul Wood, *Thomas Reid on Mathematics and Natural Philosophy* (Edinburgh University Press, 2017).
[31] Tweddle, 'An early manuscript'.
[32] Ian Tweddle ('An early manuscript' p. 17) interprets this as the abbreviation for the Latin feminine singular *Propositio altera* or 'A Second Exposition'.
[33] Grabiner, 'Newton, Maclaurin', p843.

## 4. The intellectual environment

Throughout his life, MacLaurin considered himself a devout Christian, believing that mathematics both complemented and supported his religious convictions. As he wrote circa 1734-35:

> I am satisfied that the interests of true Science and true Religion are united, & that they do real prejudice to Mankind who endeavor to represent them as opposite in any measure.[34]

The morality presented in *De Viribus* seems a natural extension of the 'physico-theology' movement that was popular in the mid-seventeenth to mid-eighteenth centuries across Britain and Europe. Physico-theology attempted 'a way of reconciling Christianity […] with the numerous scientific positions that began to prevail after 1650'.[35] English philosophers, clergymen and naturalists such as William Whiston, William Derham and John Ray published treatises seeking to establish the *existence* of a Deity from support they found in the natural world.

This was also the approach of Maclaurin's regent at Glasgow, Gershom Carmichael, in his *Synopsis Theologiæ Naturalis*.[36] Carmichael argued that Newton's laws necessitated the existence of a benevolent but immaterial being to direct and sustain the material world.[37] He taught Maclaurin that, 'the mind is a substance quite different from matter […] the mind is immediately created by God alone, and is not to be thought of as having a prior existence before its union with the body.'[38]

MacLaurin's graduation thesis on gravity followed Carmichael's views, and at the end he arrived at corollaries expressing ideas he was later to develop in *De Viribus*. Like Carmichael, he rejected the material nature of the mind and held that it could exist only in union with the body:

> I. The simple and unordered nature of the mind does not allow it to exist in any part of space in such a way that it is coextensive with it; nor indeed does it

---

[34] Colin Maclaurin, 'Letter, c. 1734-35', Aberdeen University Museums and Special Collections (A.U.) MS 206.65. This may have been composed as a letter responding to Bishop George Berkeley's 1734 critique of Newton's *Principia*, or alternatively as a draft of MacLaurin's foreword for his defense of Newton's calculus, *A Treatise of Fluxions*.
[35] Kaspar von Greyerz and Ann Blair, *Physico-Theology: Religion and Science in Europe, 1650–1750* (Johns Hopkins University Press, 2020), p2.
[36] Gershom Carmichael, *Synopsis Theologiæ Naturalis* (Edinburgh: J. Paton, 1729).
[37] Waymack, 'Moral philosophy and Newtonianism', pp.17-18.
[38] Shepherd, 'Philosophy and science…' p.132.

prevent it from being present in one place, namely, where the body is, in such a way that it is not present similarly in another place,[39]

and hinted at a motivation for *de Viribus*:

> II. Although the real or absolute essences of substances are unknown to us, it in no way follows from this that we can pronounce nothing certain concerning their dispositions and mutual relationships.[40]

MacLaurin may possibly have been influenced by his fellow Scotsman John Craig (c. 1663-1731),[41] who held that,

> It seems absurd not to be able to extend the usefulness of mathematics, 'the divine science,' beyond the narrow boundaries of this life. As the whole world of nature is made stable by geometric laws, how can anyone doubt that these lead us on to the knowledge of nature's omniscient Creator? [42]

Although, later, the two became friends,[43] it seems unlikely that they knew each other personally in 1714, as Craig had been based firmly in England since 1689. But MacLaurin might have encountered Craig's 1699 *Theologiæ Christianæ Principia Mathematica*, although there is no copy in Glasgow University Library.[44] The second half of Craig's essay, computed the 'pleasure' of the afterlife in a similar way to *De Viribus*, although it was mathematically less sophisticated.[45]

For Maclaurin, and his potential patron Colin Campbell, the development of such arguments had to take place within the restrictive 'Westminster Confession of Faith' adopted by the Scottish Parliament in 1649, which expressed Calvinist doctrine: all things were governed by God's eternal decrees, the total depravity of humanity — that people are oriented towards Satan; individuals can only be saved by God's irresistible grace, and some are predestined before birth for such grace and to be saved, while the rest are destined to remain sinners and to eternal damnation ('unconditional election'). The elected are endowed with 'living spirit' and will naturally be good because they have been elected.[46] Restrictions on universities followed doctrine and in 1699 a Parliamentary Commission listed three 'false and pernicious' propositions that regents were to guard against among students: '1. The material world has existed from eternity, 2. Our reason or philosophy is the father of Scripture; it is the criterion according to which we judge the truth of things divine, 3. A wise man's reason is the rule for morality'.[47]

---

[39] Tweddle, *MacLaurin's Physical Dissertations*, p. 25,
[40] Tweddle, *MacLaurin's Physical Dissertations*, p. 25,
[41] Andrew I. Dale, 'Craig, John (c. 1663-1731)', in *Oxford Dictionary of National Biography*, 2004.
[42] Richard Nash, *John Craige's Mathematical Principles of Christian Theology*, (Southern Illinois University Press, 1991). p. 53.
[43] Mills, *Collected Letters,* [Letter 117].
[44] Information supplied by Robert Maclean, Glasgow University Library.
[45] Nash, *John Craige's Mathematical Principles*.
[46] Brown, 'Religion and Society'.
[47] Quoted in Shepherd, 'Philosophy and science', p.305.

The penalties for transgressing these rules had been severe; in 1697 an Edinburgh student, Thomas Aikenhead, had been executed for blasphemy. Things had relaxed somewhat by 1714 as the English forced a degree of tolerance on the Scots following the 1707 Act of Union,[48] but caution was still advisable. One of Maclaurin's contemporaries was 'extruded' (expelled) from Glasgow University in 1713 for denying the doctrine of predestination,[49] while John Simson, the Enlightenment-thinking Professor of Divinity and Robert Simson's uncle, was obliged to defend his Calvinist orthodoxy before the General Assembly of the Church of Scotland in 1715 when accused of spreading false doctrine.[50]

The main charge against John Simson was that he considered human reasoning to be on a par with scriptural proclamation, thus contravening the three pernicious propositions proscribed by the Parliamentary Commission of 1699.[51] Parallelling the absolute authority of scripture over human reason, was the dogma, handed down from Augustine and Aquinas, and adopted by the early Church of Scotland, that Scriptural passages bore the absolute power of certainty and carried more weight than observation of the natural world or scientific experiment. Even in England, John Craig had come under heavy criticism for his model of the decay and corruption over time of knowledge of Christ. His opponents argued that it was impossible for divine authority, in the form of Scripture, to get corrupted in the same way that human testimony did.[52]

In 1711, the General Assembly of the Church of Scotland decreed that no one could enter the ministry unless they subscribed to the 'whole doctrine' of the Westminster Confession.[53] If he hoped for a career as a minister, MacLaurin needed to be perceived as conforming to doctrine, whatever his intellectual reservations.

Unconditional election, and the ensuing belief that no action during one's lifetime could affect one's fate after death, negating any possibility of free will, was a particularly difficult doctrine for scholars at the dawn of the Scottish Enlightenment.[54] Indeed one of the revolutionary new cornerstones of early eighteenth-century thought in Scotland was 'a distinctive emphasis on the importance of virtuous practice over orthodoxy in doctrine'.[55] MacLaurin probably inherited liberal views from his uncle Daniel, Gershom Carmichael, John Simson, and from his brother John. John was a member of the 'moderate' more rationalist wing of the Church of Scotland.[56] Thirty years later, Colin deemed John 'over Orthodox',[57] but it is not clear how close the brothers' views were in 1714 when Colin was composing *De Viribus.*

---

[48] Ahnert, *Moral Culture*; Shepherd, 'Philosophy and science', p. 310.
[49] Roger L. Emerson, *Academic Patronage in the Scottish Enlightenment* (Edinburgh University Press, 2008), p. 78.
[50] Anne Skoczylas, *Mr. Simson's Knotty Case: Divinity, Politics, and Due Process in Early Eighteenth-Century Scotland* (McGill-Queen's, 2001).
[51] Skoczylas, *Mr Simson's Knotty Case*, pp. 103-113.
[52] David Horowitz, 'Mathematics and Eschatology: English Apocalyptic Thought in the Seventeenth and Eighteenth Centuries' (University of St. Andrews MLitt thesis, 2018).
[53] Quoted in Ahnert, *Moral Culture*, p. 26.
[54] Ahnert, *Moral Culture*, esp. pp. 27-30.
[55] Christian Maurer, 'Human nature, the passions and the Fall: themes from seventeenth-century Scottish moral philosophy', in *Scottish Philosophy in the Seventeenth Century*, ed. By Alexander Broadie, (Oxford University Press, 2020), pp. 174-190, on p. 175.

To interpret the commentaries dotted throughout the essay, we must connect them to the underlying mathematics that they represent, and determine what theological principles might motivate their inclusion.

## 5. Interpreting *De Viribus Mentium Bonipetis*

MacLaurin began *De Viribus* without any preamble, by asserting that our minds seek every small piece (*particula*) of goodness in any good thing. He proceeded to examine various mathematical distributions of the good things over time, calculating the total good using fluxional calculus and offering mathematical and moral consequences for each distribution. Next, he investigated various possible microscopic force laws between the *particula* and the mind, once again exploring the consequences. Ultimately, he settled upon a force law that best reflected his own beliefs and brought his manuscript to an abrupt close without any further conclusions.

MacLaurin never provided an explanation of the 'good' that the forces of the mind are seeking in *De Viribus*. However, an interpretation that this 'good' refers to God's grace in the afterlife seems to make sense of most of the mathematical and moral observations he makes in the manuscript.[58] Indeed MacLaurin wrote on f. 5$^v$ '*dum bona vitae futurae consideramus*' ('while we are considering the good things of a future life') (fol. 5v, Tweddle p. 9,). Later in the essay he refers explicitly to 'present' good things which appear to be good things during life.

He pointed out once, early on, that his analysis applied also to 'bad things': '*Not: haec malis facile applicari posse atque alia etiam quæ sequuntur*' ('Note: these ideas and also others which follow can easily be applied to bad things') (fol. 2$^v$, Tweddle, p. 4). Elsewhere he talked of moral vices pursued by man during his lifetime, and of the temptations of Satan. The tension between good (posthumous) and bad (mortal) things sheds some light on MacLaurin's discussions of moral conflicts found later in the manuscript.

MacLaurin's model of the good-seeking forces of minds was presented at the outset of *De Viribus*. When describing the microscopic force of attraction of gravity, Newton had considered a distant point mass separated by an intervening spatial length. MacLaurin theorized a good-seeking force of minds (*vis mentium bonipeta*) of an anticipated Divine

---

[56] Henry Reay Sefton, 'The early development of moderatism in the Church of Scotland.' (Glasgow University Ph.D. thesis, 1962) pp. 2-3.
[57] Mills, *Collected Letters,* [Letter 86].
[58] Previous authors have not spotted this interpretation: Sageng, 'Colin MacLaurin' PhD. thesis; Grabiner, 'Newton, MacLaurin'; Tweddle, 'An early manuscript'; Bruneau, *Colin Maclaurin ou L'obstination Mathématicienne d'un Newtonien.*

grace separated from the present by an intervening period of time. His use of the phrase 'distance of time' (*temporis distantia*) for this period made the analogy with gravitation clear (fol. 4ʳ, Tweddle p. 6). MacLaurin hypothesized two quantities varying with time that each contributed to this force:

$I$ = the intensity of a good thing (*intensio*) at time $t$, implicitly analogous to the density of a mass at point $x$,

and

$v$ = the good-seeking accelerating force (*vis bonipetis acceleratrix*) which he defined later as varying with $t$, in implicit analogy to Newton's acceleration of gravity which varies with $x$ (fol. 4ʳ, Tweddle p. 7).[59]

Although both were measured from a present time $t = 0$, it was the posthumous time that contributed to the good-seeking force of minds that MacLaurin described; death ($t_1$ in Figure 1) marked the near boundary of the good thing. Figure 1b demonstrates the parallel.

MacLaurin began by assuming $v$ was constant, and considered $I$ alone. He included a diagram which plotted the intensity of the good thing against time (Figure 1a) and used fluxional notation to derive the total good, $\mathcal{B}$, in terms of $I$ and $t$:

$$\mathcal{B} = \mathcal{F}I\dot{t}$$

where $\mathcal{F}$ is the fluent and $\dot{t}$ is the fluxion (differential in modern terms) of time. In modern integral notation,

$$\mathcal{B} = \int_{t_1}^{t_2} I(t)\, dt$$

where $t_1 \leq t \leq t_2$ is a posthumous time period (that may be infinite).

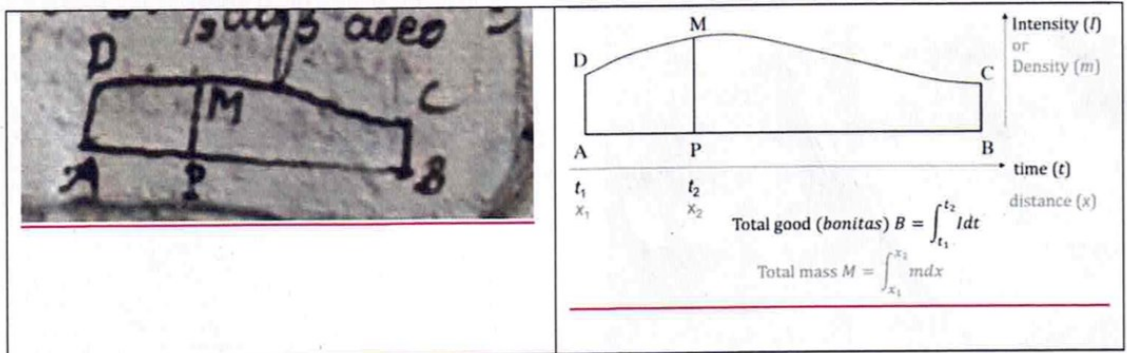

Figure 1. a) MacLaurin's diagram in *De Viribus* showing how to calculate the total good (fol. 2ʳ); b) the diagram in modern form showing the analogy with calculating a (one dimensional) total gravitational mass; the derivation of the total good is shown in black typeface, that of the mass in grey.

---

[59] John Roche points out that in the early eighteenth century, the idea of a 'force' was amorphous. Concepts such as the 'impressed force', the 'accelerative force', and the 'living force' referred to quantities that would be incommensurate by modern standards. MacLaurin's 'good-seeking accelerating force' followed in this tradition. John Roche, *The Mathematics of Measurement* (Athlone Press, 1998) pp. 99-100.

MacLaurin next discussed the total good $\mathcal{B}$ arising from various distributions of the intensity $I(t)$ (Cases A, B, C, D in Table 1) and their implications.

MacLaurin's first examples — a positioning that indicates their supreme importance — were curves of infinite length that bound a finite area with their asymptotes. Among them were the cissoid of Diocles as well as higher-order hyperbolas.[60] He did not explain why these examples were important, but they offered MacLaurin the possibility of an infinite afterlife, remaining faithful to the Apostles' Creed that promised 'life everlasting' and which was used in all Christian liturgies (including Scottish Presbyterianism) since the eighth century.[61] The reason why he required a finite total good $\mathcal{B}$, a finite total accelerating force $\mathcal{V}$,[62] and a finite total good-seeking force of minds $\mathcal{F}$, might be explained by his later discussions of the struggle between good and bad; everlasting damnation would always outweigh everlasting `bliss' if both totals were infinite.

The first such example that MacLaurin expanded upon was an intensity that 'decreases in a geometric progression while the times increase in an arithmetic progression' (Tweddle p. 6; we would say exponentially decreasing; A in Table 1). He gave a diagram (Figure 2) and used a geometric method to find the (finite) area between the curve and its asymptote:

> *Area infinita ADCB = ATED, supp. quod DT tangit curvam in A. Bonum igitur hujusmodi infinitæ durationis AB æquale est uniformi durationis AT, intensionis AD constantis.* (fol. 3ʳ)
>
> (The infinite area ADCB = ATED, where it is supposed that DT touches the curve at A. Therefore a good thing of this type of infinite duration AB is equal to a uniform good thing of duration AT and of constant intensity AD.) (Tweddle p. 6)

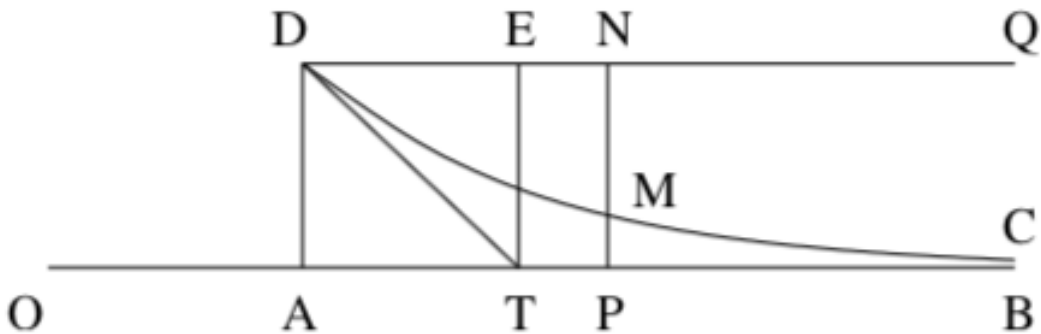

---

[60] The cissoid of Diocles is a cubic curve, important for constructing mean proportionals, given in modern polar coordinates by $r = 2a \sin\theta \tan\theta$.
[61] Jan Milič Lochman, 'Apostles' Creed', in *The Encyclopedia of Christianity*, ed. by Erwin Fahlbusch et al, 5 vols, (Eerdmans, 1999), I pp. 109-110.
[62] $\mathcal{V}$ and $\mathcal{F}$ are our notation; MacLaurin did not assign symbols to the total accelerating force or total good-seeking force of minds.

Figure 2. Redrawing of MacLaurin's diagram (fol. 4ᵛ). The unbounded area between the curve DMC and its asymptote OAB is equal to the area of the rectangle ATED where DT is tangent to the curve at point D.⁶³

MacLaurin presented this method for finding the unbounded area without proof, suggesting that it was well known. In fact, it can be traced to Torricelli.⁶⁴ MacLaurin may have learnt it as a student through Christian Huygens' 'Discours de la cause de la pesanteur' which cites Torricelli; Robert Simson owned a copy of Huygens' work.⁶⁵

MacLaurin's next extended example was of a constant intensity $I = k$ (D in Table 1). Supported by an explanatory diagram (Figure 3) he observed:

> *Infinité magis præstat hujusmodi infinité durantis boni intensionem vel maximé exigua quantitate augere quam ejus durationem quam maximé prolongare; Nam addendo DL intensioni AD boni infiniti ADCB, augmentum infinitum DLOC ei adjungimus; addendo autem AQ durationi infinita AB fit incrementum finiti Boni AQHD.* (fol. 3ᵛ)

> (It is infinitely more preferable to increase the intensity of an infinitely lasting good thing of this type by the smallest possible quantity than to extend its duration as far as possible; for by adding DL to the intensity AD of the infinite good thing ADCB, we adjoin the infinite increase DLOC to it; but by adding AQ to the infinite duration AB the increment of the finite good thing AQHD results.) (Tweddle p. 6)

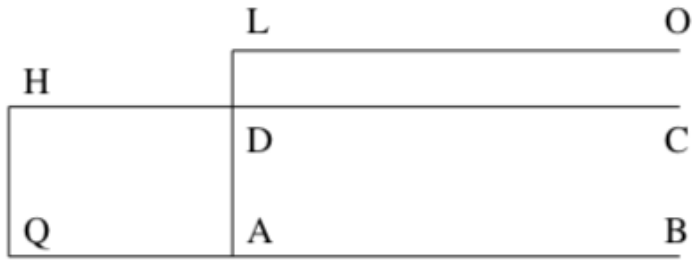

Figure 3. Redrawing of MacLaurin's diagram (fol. 3ᵛ).

---

⁶³ In modern terminology, the intensity $I = I_0 e^{-kt}$ is represented by the curve DMC. Then for any time $t = a$, the point D has coordinates $(a, I_0 e^{-ka})$. Define a point $T = (a + \frac{1}{k}, 0)$, the intersection of the horizontal axis and the tangent to DMC at D. It follows that the unbounded area beneath the curve is finite:
$$\text{ADCB} = \int_a^\infty I_0 e^{-kt}\, dt = \frac{1}{k} \times I_0 e^{-ka} = \text{AT} \times \text{AD} = \text{ATED}.$$
⁶⁴ Gino Loria, 'Le ricerche inedite di Evangelista Torricelli sopra la curva logarithmica', *Bibliotheca Mathematica: Zeitschrift für Geschichte der Mathematischen Wissenschaften,* 3ʳᵈ ser., 1 (1900), pp. 75-89; Lorenzo J.,Curtis, 'Concept of the exponential law prior to 1900', *American Journal of Physics*, 46 (1978), pp. 896-906.
⁶⁵ Huygens, Christiaan, 'Discours de la cause de la pesanteur', in *Traité de la Lumiere* (Leiden: Pierre vander Aa, 1690), pp. 129-180; G.U.L. Sp Coll Ea5-d.8.

MacLaurin was stating that joining the finite area AQHD to the infinite region ADBC has less effect than appending the infinite strip DLOC (regardless of how small the positive increment DL might be). His assertion was typical of contemporary treatments of the infinite, though it is incommensurable with the modern understanding of infinity for which the inclusion of either area would have no effect on the resulting total.

Then comes perhaps the most difficult passage in *De Viribus* both to translate and to understand:

> *Bonorum igitur virorum de miseriis hujus vitæ querelas debent, perimere consideratio incrementi quod exinde intensioni fœlicitatis futuræ æternæ accedit, quo (forte) fit ut eorum fœlicitas tota simul sumpta major sit quam si ab initio eorum existentiæ exordium sumpsisset, & nunquam cecidisset homo. Intensionem autem augeri ex eo satis patere videtur quod major sit Gratitudo & quod quamplurimæ virtutes exerceantur quibus vix ullus esset locus si omnes homines Innocentes sine ullis malis aut infortuniis vitam agerent.* (fol. 3$^v$-4$^r$)

> (Therefore the complaints of good men about the miseries of this life must be removed by the consideration of the increment which is added as a result to the intensity of future eternal happiness, by which it comes about (by chance) that their total happiness altogether is greater than if it had started from the beginning of their existence and man had never fallen. Moreover it seems to be quite clear that the intensity is increased on account of the fact that gratitude is greater and very many virtues are exercised for which there would be scarcely any place if all innocent men were to conduct their life without any bad things or misfortunes.) (Tweddle p. 7)

MacLaurin seems to be arguing that due to the Fall (resulting from original sin), and the consequent miseries of this life, man's baseline happiness (HDC in Figure 3) was lower than had there been no Fall — the difference between the baseline and the good of the afterlife was correspondingly greater, equivalent to an increment, and their total happiness throughout time would thus be greater.

John Craig had offered a similar, though less developed, thought in 1699:

> The value of the pleasure promised by Christ is infinitely greater than the value of the pleasure of our present life. For the pleasure promised by Christ is of nondecreasing intensity and of infinite duration [..,] but the pleasure of our present life is of finite intensity and also of finite duration.[66]

MacLaurin's next sentence, beginning, 'Moreover it seems […]' can be interpreted in two ways:

---

[66] Nash, *John Craige's Mathematical Principles*, p. 81.

1. MacLaurin could be suggesting a secondary effect — that gratitude induced by the perception of greater good to come led men to do good deeds that would have been unnecessary if the Fall had not happened, further incrementing the reward after death. Such a suggestion, that actions in this life might affect one's reward in heaven in any way, even to the extent of an increment, may be seen as modifying the doctrine of unconditional election, but would be in line with the emphasis on virtuous practice characteristic of liberal Presbyterian scholars at the time.[67] MacLaurin's approach to God's grace in this passage would, then, be a further indication of the extent of the evolution of Church of Scotland theology by 1714.
2. Alternatively, MacLaurin may be adducing 'the fact that gratitude is greater and […] virtues are exercised' as evidence that the intensity of the good had, indeed, been incremented by the Fall, as he had suggested in the first half of his paragraph. Sageng's alternative translation of this sentence makes this interpretation more likely:

   > That the intensity is increased seems to be sufficiently clear from the fact that gratitude will be greater and that very many virtues will be exercised for which there would be scarcely any place if all men lived their lives as innocents without any evils or misfortunes.[68]

Either way, MacLaurin's reflections were paralleled sometime later by Francis Hutcheson:

> The mistakes, imperfections, provocations, calumnies, injuries, or ingratitude of others we shall look upon as matters presented to us by providence for the exercise of the virtues God has endued us with, by which we may more approve our selves [sic] to his penetrating eye, and to the inward sense of our own hearts, than by the easier offices of virtue where it has nothing to discourage or oppose it.[69]

It is not clear whether Hutcheson knew of *De Viribus*. Similarities to MacLaurin in his thinking are not surprising as they were contemporaries as students at Glasgow and both attended John Simson's divinity lectures.

Having discussed these distributions of the intensity of a good thing, MacLaurin introduced the possibility that the good-seeking accelerating force *v* might also vary with 'distance of time' from the observer (the living person) whose mind experienced the force. After introducing *v*, MacLaurin seldom used algebraic or fluxional notation and expressed his reasoning in words and geometrical diagrams in the rest of the essay.

---

[67] Maurer, 'Human nature' p. 175.
[68] Sageng, 'Colin MacLaurin' PhD. thesis, p. 125.
[69] Published posthumously in 1755, but composed and privately circulated before 1738, and based on public lectures given between 1730 and 1738: Frances Hutcheson, *A System of Moral Philosophy, in Three Books*, 2 vols, (Glasgow: R. and A. Foulis, 1755), I; Luigi Turco, 'Introduction' in Francis Hutcheson, *Philosophiae Moralis Institutio Compendiaria, with A Short Introduction to Moral Philosophy* (Indianapolis: Liberty Fund, 2007), pp. ix-xxiii.

Figure 4 shows one example of MacLaurin's description of how to combine the intensity of a good thing and the force law.

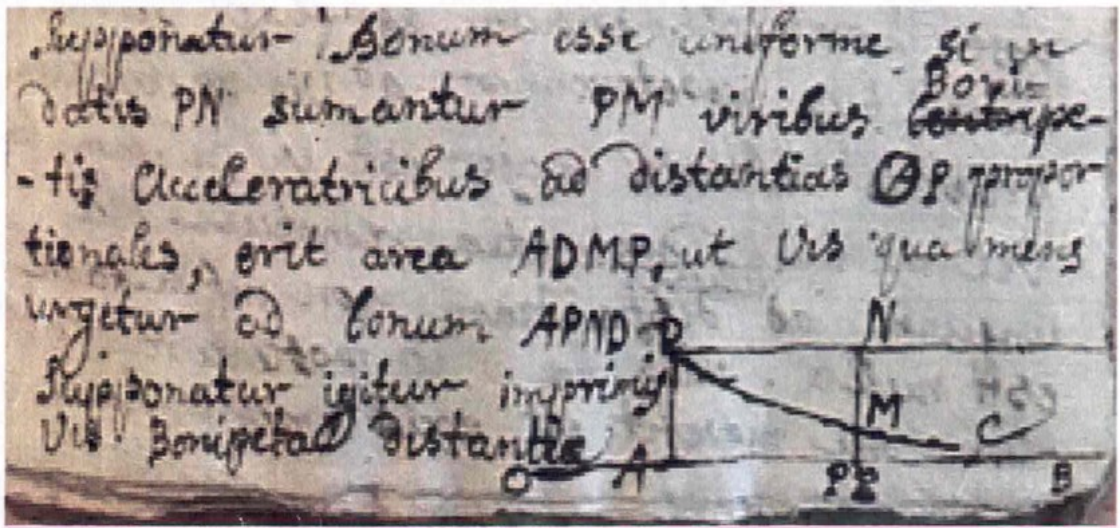

Figure 4. 'Let it be supposed that the good thing is uniform. If in the given PN the PM are taken proportional to the good-seeking accelerating forces at the distances OP, the area ADMP will be as the force by which the mind is driven towards the good thing APND.' (fol. 4$^r$, Tweddle pp. 7-8).

MacLaurin's diagram, which shows the curve representing the force law DMC terminating at the vertex D, has been misinterpreted by commentators as taking A (rather than O) as the origin of time.[70] However, MacLaurin's diagram clearly shows O, and he explicitly states that it is the 'distance of time' OP that defines the good-seeking accelerating force $v$. In Figure 5, which expresses MacLaurin's reasoning in modern form, we avoid this confusion by showing the force law $v$ (dotted) as clearly distinct from the intensity $I$ of the generalized posthumous good pictured in Figure 1 with which the force is being combined.

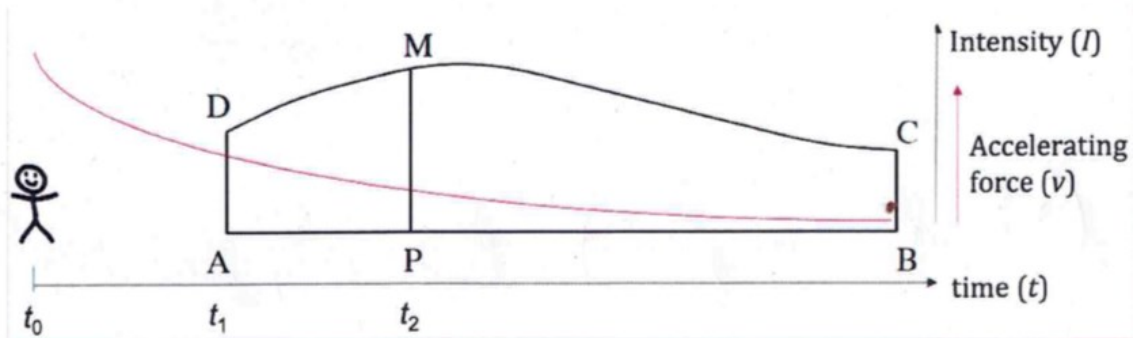

---

[70] Tweddle acknowledges that MacLaurin's 'distance of time' is 'just the elapsed time from the start' but does not explicitly specify when this 'start' occurs. Sageng does not mention 'distance of time'. Tweddle,'An early manuscript' p. 13; Sageng,'Colin Maclaurin' PhD. thesis.

Figure 5. The intensity $I$ of a good thing (solid curve) and the good-seeking accelerating force $v$ (dotted curve) plotted on a posthumous time period $t_1 \leq t \leq t_2$.

Thus, in our modern form,

Good-Seeking Force of Minds, $\mathcal{F} = \int_{t_1}^{t_2} I(t) \cdot v(t)\, dt$

measures the total good-seeking force exerted on the living mind by the posthumous time period $t_1 \leq t \leq t_2$.

From this point, MacLaurin adhered to a generalized Newtonian method but did not adopt the inverse square form of gravitation. Instead, he held the intensity $I$ constant (except in one case) while examining various forms for $v$. For each, he commented on the resulting $\mathcal{F}$ from a mathematical perspective and, in line with Calvinist doctrine, tested his results against theology rather than the natural or observed social world. Table 1 shows the various functional forms of $I$ and $v$ considered.

Table 1. Forms of Quantities mentioned in *De Viribus* in modern notation (letters in the first column are the authors' and used solely for reference).

| Our Reference (in the order presented in *De Viribus*) | $I(t)$ **Intensity of a Good Thing, with Tweddle's translation of MacLaurin's description** | $v(t)$ **Good-Seeking Accelerating Force, with Tweddle's translation of MacLaurin's description** |
|---|---|---|
| A | $I = I_0 a^{-kt}$<br><br>'the intensity decreases in a geometric progression while the times increase in an arithmetic progression […] the common logarithm' | $v = 1$<br><br>[$v$ not stated but later described as assumed constant] |
| B | $I = k(t - t_1)$<br><br>'the intensities of the good thing increase uniformly taking their beginning from nothing' | $v = 1$<br><br>[$v$ not stated but later described as assumed constant] |
| C | $I = k(t - t_1)^\beta \quad (\beta > 0)$<br><br>'the intensity is as some positive power of the duration passed through' | $v = 1$<br><br>[$v$ not stated but later described as assumed constant] |
| D | $I = k$<br><br>'uniform good things whose intensities are variable' | $v = 1$<br><br>[$v$ not stated but later described as assumed constant] |
| E | $I = 1$<br><br>'the good thing is uniform' | $v = \dfrac{k}{t}$<br><br>'let the good-seeking force be supposed to be reciprocally proportional to the distance of the time' |
| F | $I = k_1 t$<br><br>'good things which are as the distances of the times of the mind from them' | $v = \dfrac{k_2}{t}$<br><br>'let the good-seeking force be supposed to be reciprocally proportional to the distance of time' |
| G | $I = 1$<br><br>[$I$ not stated but assumed constant] | $v = k a^{-\beta t} \quad (\beta > 0)$<br><br>'the good-seeking accelerating forces decrease in a geometric progression if the times increase in an arithmetic progression' |
| H | $I = 1$<br><br>[$I$ not stated but assumed constant] | $v = \dfrac{\left(2\sqrt{a} - \dfrac{t}{\sqrt{a}}\right)^2}{4}$<br><br>$(0 \leq t \leq 2a)$<br><br>'let $\dot{v}$ be as $v^{1/2}$ or $\sqrt{v}$' |

At least one of the two quantities *I* and *v* was held constant in all but case F, allowing MacLaurin to consider the total good-seeking force of minds to be proportional to either the total goodness (*bonitas*) of God's grace

$$\mathcal{B} \propto \int I \, dt$$

or to the total accelerating force (*vis*) on the mind which we (but not MacLaurin) denote by $\mathcal{V}$:

$$\mathcal{V} \propto \int v \, dt$$

given here in modern (integral) notation. It seems likely that MacLaurin did not know, in general, how to integrate the result in cases where both varied.

The first varying force that MacLaurin proposed was one 'reciprocally proportional to the distance of time' (i.e. a simple hyperbolic force). He discussed four corollaries. Corollary 3 (F in Table 1), was that this force, if combined with an intensity varying in direct proportion to the distance of time, gave a constant total force whenever it was observed:

> 3: *Secundum hanc hypothesin bona quae sunt ut distantiae temporum mentis ab iis aequaliter appetuntur.* (fol. 4ᵛ)
>
> (According to this hypothesis [of inversely proportional $v$] good things which are as the distances of the times of the mind from them, are equally sought.) (Tweddle p. 8).

In modern terms, $I.v = k_1 t . k_2/t = k_1 k_2$, a constant.

In corollary 4, MacLaurin moved to considering 'present' good things by which he appears to mean imminent things in mortal life, before death. He pointed out that as $t$ approached zero (i.e. the present time), then the good-seeking force rose rapidly towards infinity:

> 4: *Vis in bonum præsens infinité major est vi in idem bonum vel minima distantia remotum. Nec unquam ad ullam distantiam non infinitam est nulla.* (fol. 4ᵛ)
>
> (The force into the present good thing is infinitely greater than the force into the same good thing removed even by a very small distance. And never at any non-infinite distance is it nothing.) (Tweddle p. 8)

Whereas the modern mathematician would say that $v$ was undefined at $t = 0$, MacLaurin interpreted it as infinite. He was deeply disturbed by this conclusion:

> *Atque haec duo posteriora Corollaria me movent ut existimem hanc hyp. non esse veram* (fols. 4ᵛ-5ʳ)
>
> (And these last Corollaries [3° and 4°] trouble me to the extent that I am of the opinion that this hypothesis is not true.) (Tweddle p. 7)

What bothered MacLaurin so much about corollaries 3° and 4° that he discarded a reciprocally decreasing (i.e. hyperbolic) force? We can only conjecture.

Perhaps it was finding 'infinite' moral quantities as consequences of his mathematical analyses. Corollary 3° may have prompted MacLaurin to reject Case F since the constant $k_1 k_2 \neq 0$ meant that he could not compute the total good-seeking force of minds over the unbounded posthumous time beyond $A$ necessitated by the Apostles' Creed. Similarly, the 'infinite' nature of $(I \cdot v)|_{t=0}$ in Corollary 4° implied an infinite present force into the good. This would be beyond mortal experience, and would deny any possibility of free will, running contrary to MacLaurin's probably liberal views. Carmichael's teaching had addressed the relation between God, the spirit, free will, and the mind, and acceptance of free will as tempering the harshness of unconditional election was growing among progressive Presbyterians .[71] Alternatively, MacLaurin may have been interpreting Corollary 4° in the context of 'bad things': an infinite present force towards Satan would overwhelm any present good-seeking force and preclude all mortal human virtue. His discussion of his next force, a geometrically decreasing one, suggests that consideration of bad things may, indeed, have been what bothered MacLaurin.

In his analysis of a geometrically decreasing force (i.e. exponentially decreasing — G in Table 1), MacLaurin referred to Newton's demonstration that, '*decrementa virium sive uxiones erunt ut ipsae vires*' (fol. 5$^{rr}$) ('the decreases or fluxions of the forces will be as the forces themselves') (Tweddle p. 9).This is the result that we know as the differential (here the rate of decrease) of an exponential function is proportional to the exponential function itself; so the greater the force, the faster it decreases. The implication of this result was that:

> *Quae hyp. malitiæ ac diligentiæ Temptatoris est satis consentanea, qui verisimiliter majorem adhibet operam ad vim cohibendam quo major est*. (fol. 5$^r$)
>
> (This hypothesis is quite in accordance with the malice and diligence of the Tempter, who probably applies greater effort to the containment of force the greater it is.) (Tweddle p. 8)

MacLaurin seems to be suggesting that a greater good-seeking force elicits a greater effort by the Tempter (i.e. the devil or Satan), to diminish that force. This was consistent with Christian beliefs about the devil: in the biblical narrative of the three temptations of Christ (Luke 4:1-13, Matthew 4:1-11), the lure of each temptation becomes stronger with each of Jesus's rejections. Language of the 'malice' and 'diligence' of the Tempter was current around the turn of the seventeenth/eighteenth century. The English Presbyterian, William Bates (1625-1699), for example, wrote that the Tempter's 'diligence is equal to his malice'.[72]

MacLaurin's treatment of a geometrically decreasing force applied the same quadrature method used earlier for a geometrically decreasing intensity. He observed that the good-seeking accelerating force *v* extending over an unbounded posthumous time interval produced only a finite total natural acceleration $\mathcal{V}$. Consequently, he suggested, for some this might easily be outweighed by the attraction of vices (bad things) in this lifetime;

> *Atque hinc prodiit modus solvendi phænomenon insigne: $\overline{num}$ quod plurimi qui virtutem bonum infinitum post se trahere opinare videntur, vitia tamen sequuntur.* (fol. 5$^v$)

---

[71] Shepherd, 'Philosophy and science'; Alasdair Raffe, 'Presbyterians and Episcopalians: the formation of confessional cultures in Scotland, 1660-1715', *English Historical Review* 514 (2010), pp. 570-598.
[72] W. Farmer, *The whole works of the Rev. W. Bates, D. D.,* 4 vols, (London: James Black, 1815), II p. 122.

(And hence comes forth a means for explaining a remarkable phenomenon: namely, that very many people who seem to be of the opinion that virtue brings an infinite good thing after itself, nevertheless pursue vices.) (Tweddle p. 8)

He generally liked the hypothesis of a geometrically decreasing force, but it had one difficulty — that the good-seeking accelerating force, $v$, vanishes only at infinite distance. Whether this would be an acceptable force for the good of the afterlife MacLaurin does not say, but he is explicit that '*in bonis hujus vitae obtinere non posse videtur*' (fol. 6$^r$), ('it cannot hold in the good things of this life') (Tweddle p. 10). He may have felt that the time-limited things of this life required a time-limited force law. His comment demonstrates his concern to develop a *universal* law that would apply to all good things whatever their distance of time, analogous the Newton's gravitation that applied to apples on earth and to planetary systems.

Seeking such a force became Case H, MacLaurin's final form. To achieve this, he first ensured that the force at the present time was finite, $v(0) = a$ for a fixed value $a > 0$, after which he suggested:

> *Videamus igitur quam aliam vis bonipetæ potentiam eligamus cui proportionalem ejus decrementum statuamus.* (fol. 6$^r$)

> (Let us see what other power of the good-seeking force we may select to which we set its decrease proportional.) (Tweddle p. 10)

In modern notation, Maclaurin wished the good-seeking accelerating force $v$ to satisfy
$$\frac{dv}{dt} = -cv^m$$
for some power $m$ (where $c > 0$ is a fixed constant of proportionality). He proceeded to give his reasons for ruling out values where $m \leq 0$ (since $v$ would increase and thus not vanish at any finite time), $m = 1$ (since this would reduce to Case G which he had already discarded), and $m > 1$ (since this would result in an infinite good-seeking accelerating force into some present good thing which was dismissed in Case F). Thus, MacLaurin was left to pick $m$ such that $0 < m < 1$; he arbitrarily chose $m = 1/2$ because it was the mean of 0 and 1 (fol. 6$^v$, Tweddle p. 10), which resulted in a quadratic solution,
$$v(t) = \frac{\left(2\sqrt{a} - \frac{t}{\sqrt{a}}\right)^2}{4}$$
for $0 \leq t \leq 2a$, a parabolic curve that has its vertex and vanishes at $t = 2a$ (point B in Figure 6).

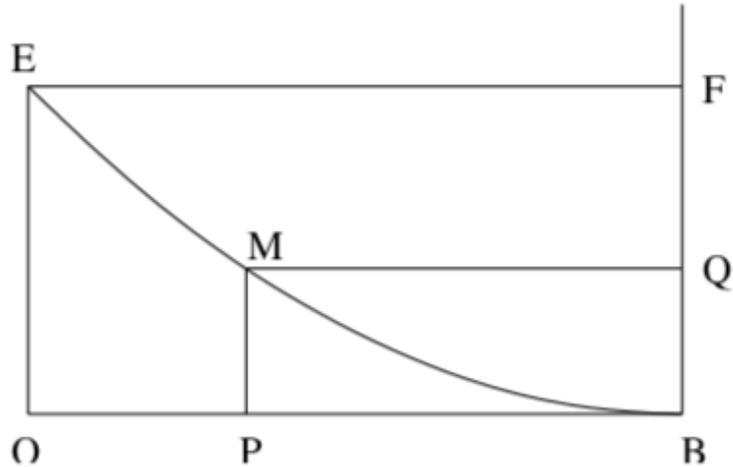

Figure 6. Redrawing of MacLaurin's diagram for case H (fol. 6ᵛ), the parabolic curve EMB with vertex B and P a point between O and B.

MacLaurin then used the geometric properties of conics to show that the force *v* satisfied his requirements:

> 1, *vis bonipeta ad distantiam finitam OB est nulla. Hinc*
> 2, *est vis bonipeta in duplicata ratione distantiae temporis a puncto in quo evanescit.* (fol. 6ᵛ)
>
> (1. The good-seeking force is zero at the finite distance OB, Hence:
>  2. The good-seeking force is in the duplicate ratio of the distance of the time from the point in which it vanishes.) (Tweddle p. 11).

Thus, he concluded his essay by stating the total good-seeking force on the mind, given by the area OEMP, as

$$\text{OEMP} = \frac{OB^3 - BP^3}{12\,OE}$$

.

## 6. Discussion

While MacLaurin was undoubtedly participating in the physico-theological movement, in *De Viribus,* he began to take an independent line. Unlike Carmichael and the English authors, he did not seek to prove the existence of God but took this for granted. Instead, he sought to understand the mechanism through which the human mind responded to God. Later in the century, Hutcheson, Hume, and Turnbull, were all to take up 'the notion of introducing the method of experimental reasoning into moral subjects, and of thereby doing for the problem of the mind what Newton had done for the problem of matter.'[73] But *De Viribus* not only demonstrates that such debates were current in Scotland much earlier, in the 1710s, but also represents a far more Newtonian and technical analysis than these other authors.

The next evidence we get of mathematising approaches to morality in Scotland is Francis Hutcheson's work in the 1720s. He and MacLaurin knew each other and in a letter to Hutcheson written in 1728, MacLaurin ruminated on 'my old acquaintance' and their commonality of thought regarding moral philosophy,[74] a commonality that may originate in their shared experiences as students at Glasgow. But even MacLaurin had antecedents among Scots, though perhaps not in Scotland. John Craig was one such, and Craig's patron, Gilbert Burnet, Bishop of Salisbury may be another. MacLaurin certainly met Craig,[75] and maybe Burnet also, when visiting London in 1718. Further research is required to illuminate the Scottish pre- and early-Enlightenment networks of discussion extending across the British Isles and Continental Europe that presaged the work of Hutcheson, Hume and Turnbull. They did not spring from nothing.

Pursuing the analogy between MacLaurin's good-seeking forces, and Newtonian gravitation, brings aspects of *De Viribus* into sharp focus. We remarked earlier that MacLaurin was clearly aiming at a universal law that would draw together all human experiences of good- and bad-seeking forces in a single mathematical formulation. Other parallels with gravitational force are more conjectural and would require significant further research, especially into Carmichael's and John Simson's teaching, to substantiate. Although MacLaurin believed that minds lacked substance, unlike gravitating objects, were they like such objects in requiring a force to change their motion (possibly construed as disposition towards good or vice)? An infinite good- or bad-seeking force would be irresistible and preclude any exercise of free will or change in disposition towards good or bad; this would be unconditional election at its most fundamental. MacLaurin's requirement of a finite force seems to allow some possibility of change. But even if finite, the good-seeking force was, like gravity which MacLaurin believed to be the 'will of some incorporeal and intelligent cause', outside of man's control.[76] The good-seeking force was determined by God or, for bad-seeking, by Satan, but perhaps man had some limited power to accede to, or resist, it?

This brings us to a major way in which MacLaurin's force differed from gravitation. Newton's law was reciprocal and mutual; each object acted on the other and the masses of both bodies were required to calculate the attraction. There is no such mutuality in the good-seeking force; only the total good is required and the mind has no effect on the good;[77] indeed, the 'good' and the mind appear to be very different types of entity. It is possible that MacLaurin deemed any force that the mind might exert to be so insignificant compared with that of the good that it could be overlooked, as in any effect an apple's gravity might have on the earth.

---

[73] George Davie, *The Scotch Metaphysics: A Century of Enlightenment in Scotland* (Routledge, 2001), p. 12.
[74] Mills, *Collected Letters*, [Letter 13]
[75] Mills, *Collected Letters,* [Letter 117].
[76] MacLaurin, *De Gravitate*, 1713, trans. and quoted in Sageng, 'Colin MacLaurin' Ph.D. thesis, p. 120.
[77] We thank Jane Wess for highlighting this asymmetry.

A more intriguing possibility, and one that might better align with the differing natures of mind and good, is that MacLaurin's 'force' was more akin to what, today, we could call 'gravitational potential', a quantity that pertains to an individual object depending only on its mass and the distance from it, that is combined with the mass of an attracted object to work out the mutual attraction.[78] Roche has pointed that ideas of force in the seventeenth and eighteenth centuries were not well differentiated and that *potentia* was used alongside *vis*.[79] It would not be surprising if MacLaurin's concept was closer to our 'potential' than to our 'force', even if he did term it *vis*. If this is the case, it would demonstrate a strong continuity with medieval scholastic concepts of *potentia* or potentiality. Stephan Schmid describes how late medieval authors moved from ascribing natural processes to a potentiality intrinsic to bodies or substances, to ascribing them to 'the rational potentialities of God, his will, or intellect'.[80] Michael-Thomas Liske and Katia Saporiti trace the ongoing influence of these views in the seventeenth and early-eighteenth centuries.[81] If MacLaurin was, indeed, mathematising such concepts, this would evidence a bridge between the scholastic *potentia* and the development of the mathematical theory of potentials by Laplace, Gauss and Poisson at the end of the eighteenth century.

As noted by Sageng, another approach that MacLaurin seems to have inherited from the scholastics is that of graphing.[82] His diagrams in *De Viribus* utilise Nicolas Oresme's graphical methods of representing qualities and motion in two dimensions, and are distinctively different from Newton's in the *Principia*.[83] Further research into student notebooks might reveal how widespread such practices were in Scotland in the early seventeenth century.

MacLaurin's treatment of the infinite is surprising in many places to the modern mathematician, but he was in line with his contemporaries. As in the case of *De Viribus,* consideration of the infinite during the seventeenth and eighteenth centuries was often in the context of religion or morality.[84] As Rob Ilife has observed, '…the comparison between performing operations involving mathematical infinities, and grasping religious infinities, was one that would reappear in [Newton's] writings'.[85]

---

[78] For gravitation, the potential $\varphi$ of an object of mass $M$ at at distance $r$ is $\varphi \propto -M/r$; the corresponding attraction $F$ on a second object mass $m$ is $F \propto Mm/r^2$.
[79] Roche, *Mathematics of measurement*, p. 99.
[80] Stephan Schmid, 'Potentialities in the Late Middle Ages—The Latin Tradition', in *Handbook of Potentiality*, ed. by Kristina Engelhard and Michael Quante (Springer Netherlands, 2018), pp. 123–53, on p. 123.
[81] Michael-Thomas Liske, 'Potentiality in Rationalism', in *Handbook of Potentiality*, ed. by Kristina Engelhard and Michael Quante (Springer Netherlands, 2018), pp. 157–97; Katia Saporiti, 'Potentiality in British Empiricism', in *Handbook of Potentiality*, ed. by Kristina Engelhard and Michael Quante (Springer Netherlands, 2018), pp. 199–226.
[82] Sageng, 'Colin MacLaurin' Ph.D. thesis, p99.
[83] See, for example, Walter Roy Laird, 'Change and Motion', in *The Cambridge History of Science*, 2 (2013), pp. 404–35.
[84] See, e.g. Ahad Nachtomy and Reed Winegar (eds), *Infinity in Early Modern Philosophy* (Cham, Switzerland: Springer, 2018).

It has been claimed that MacLaurin was one of only a handful who properly understood Newton's *Principia* in the early eighteenth century.[86] But *De Viribus* demonstrates not only Newtonianism, but continuities from the medieval to the early modern world. It shows how the Scottish culture within which MacLaurin operated affected his thought processes and motivations, opening up new opportunities for questioning interactions between mathematics and society.

## 7. Acknowledgements

Ian Tweddle kindly made his translation of *De Viribus* available to us and has been unfailingly helpful in discussing this and his interpretation of MacLaurin's mathematics. Robert Maclean, Assistant Librarian at Glasgow University Library, has painstakingly investigated all our questions about Glasgow's holdings in MacLaurin's time and we are grateful to him. Christine Shepherd's thesis on *Philosophy and Science In the Arts Curriculum of the 'Scottish Universities in the 17th Century* has been invaluable and she kindly discussed points of detail with us and set David Horowitz on the track of the missing MacLaurin notebook.

---

[85] Rob Iliffe, *Priest of Nature: The Religious Worlds of Isaac Newton* (Oxford University Press, 2017), p. 101.
[86] See, for example, Turnbull, *Bicentary of the death of Colin MacLaurin*, p.1.

Hutcheson, Francis, *A System of Moral Philosophy, in Three Books; Written by the Late Francis Hutcheson, Published from the Original Manuscript by his son Francis Hutcheson, to Which is Prefixed Some Account of the Life, Writings and Character of the Author by William Leechman*, 2 vols, (Glasgow: Printed and sold by R. and A. Foulis; London: Sold by A. Millar Over-Against Katherine-Street in the Strand, and by T. Longman in Pater-Noster Row, 1755), I

Huygens, Christiaan, 'Discours de la cause de la pesanteur', in *Traité de la Lumiere : Où Sont Expliquées les Causes de Ce Qui Luy Arrive dans la Réflexion, & dans la Refraction. Et Particulièrement dans l'Etrange Refraction du Cristal d'Islande*, (Leiden: Pierre vander Aa, 1690), pp. 129-180

Iliffe, Rob, *Priest of Nature: The Religious Worlds of Isaac Newton* (Oxford: Oxford University Press, 2017)

Liske, Michael-Thomas, 'Potentiality in Rationalism', in *Handbook of Potentiality*, ed. by Kristina Engelhard and Michael Quante (Springer Netherlands, 2018), pp. 157–97

Lochman, Jan Milič, 'Apostles' Creed', in *The Encyclopedia of Christianity*, ed. by Erwin Fahlbusch et al, 5 vols, (Eerdmans, 1999), I pp. 109-110

Loria, Gino, 'Le ricerche inedite di Evangelista Torricelli sopra la curva logaritmica', *Bibliotheca Mathematica: Zeitschrift für Geschichte der Mathematischen Wissenschaften*, 3rd ser., 1 (1900), pp. 75-89

Maas, Harro, 'Where mechanism ends: Thomas Reid on the moral and the animal oeconomy', *History of Political Economy* 35 (annual supplement 2003), pp. 338-360

MacLaurin, Colin, *Dissertatio Philosophica Inauguralis, de Gravitate, aliisque Viribus Naturalibus*, (Edinburgh: Apud Robertum Freebairn, Typographum Regium, 1713)

MacLaurin, Colin and Patrick Murdoch, *An Account of Sir Isaac Newton's Philosophical Discoveries, in Four Books. By Colin MacLaurin, A.M. Late Fellow of the Royal Society, Professor of Mathematics in the University of Edinburgh, and Secretary to the Philosophical Society there. Published from the Author's Manuscript Papers* (London: the Author's Children, 1748)

Maurer, Christian, 'Human nature, the passions and the Fall: themes from seventeenth-century Scottish moral philosophy', in *Scottish Philosophy in the Seventeenth Century*, ed. By Alexander Broadie, (Oxford University Press, 2020), pp. 174-190

McKay, Johnston R., 'The MacLaurin papers', *The College Courant, The Journal of the Glasgow University Graduates Association*, 25 (1973), pp. 30-34

Mills, Stella (ed.), *The Collected Letters of Colin MacLaurin* (Nantwich: Shiva Publishing Limited, 1982)

Moore, James and Silverthorne, Michael, 'Carmichael, Gershom (1672-1729)', in *Oxford Dictionary of National Biography*, 2003

Nachtomy, Ahad and Winegar, Reed (eds.), *Infinity in Early Modern Philosophy* (Cham, Switzerland: Springer, 2018)
28